\documentclass[slac_one]{revtex4}
\pdfoutput=1
\usepackage{graphicx}
\usepackage{xspace}
\usepackage{fancyhdr}
\newcommand{\nue}         {$\nu_{e}$\xspace}
\newcommand{\numu}        {$\nu_{\mu}$\xspace}

\newcommand{\tonethree}   {$\theta_{13}$\xspace}
\newcommand{\tonetwo}     {$\theta_{12}$\xspace}
\newcommand{\ttwothree}   {$\theta_{23}$\xspace}
\newcommand{\msqonetwo}   {$\Delta m^2_{12}$\xspace}

\newcommand{\msqtwothree} {$\Delta m^2_{23}$\xspace}

\graphicspath{{./figures/}}

\begin{document}

\title{Experimental Neutrino Physics} 

\author{Christopher W. Walter}
\affiliation{Duke University, Durham, NC 27708, USA}

\begin{abstract}

  In this talk, I review the framework that explains our experimental
  observations in neutrino physics.  I will explain how a set of
  measurements in the last few years have filled in remaining gaps in
  this picture, and layout the remaining outstanding questions along
  with techniques for addressing them in the coming decade.

\end{abstract}

\maketitle
\thispagestyle{fancy}

\section{INTRODUCTION}

It's been a remarkable decade in neutrino physics. Ten years ago this
summer, at the 1998 neutrino conference in Takayama, the
Super-Kamiokande collaboration reported the observation of neutrinos
changing flavor~\cite{fukuda:1998mi}, thereby establishing the
existence of neutrino mass.  A few years later, the SNO experiment
solved the long-standing solar neutrino problem~\cite{Ahmad:2002de}
demonstrating that it too was due to neutrino oscillation.  Just a few
years after that, these effects were confirmed and the oscillation
parameters were measured with man-made neutrino
sources~\cite{Ahn:2002up,Araki:2004mb}. Now, just in this last year,
the same neutrinos which were the source of the 30~year old solar
neutrino problem were measured for the first time in a real-time
experiment.

In this talk, I will explain how a set of experiments, especially ones
in the last few years, have established a consistent framework of
neutrino physics and also explain some outstanding questions. Finally,
I will cover how a set of upcoming experiments hope to address these
questions in the coming decade.

\subsection{Framework}

First, I would like to briefly review our current understanding of the
properties of neutrinos.  As far as we know, there are three light
neutrinos which interact via the weak force.  Since there are three
neutrinos, there are two mass splittings that separate them, and two
possible configurations or ``hierarchies'' which are presently
consistent with the experimental data.  The first, known as the
``normal'' hierarchy is where the two lowest mass states are separated
by the small solar splitting and then the highest mass state is
separated from those two by the much larger atmospheric mass
splitting.  The so called ``inverted'' hierarchy reverses this
configuration and the two highest mass states share the smaller
splitting~\cite{Yao:2006px}.

The eigenstates that travel through space are not the flavor states
that we measure through the weak force, but rather the mass
states. Since each flavor state is a quantum mechanical combination of
mass states, there is some probability to measure a different flavor
state then the original one after some time and distance.

The oscillations of atmospheric $\nu_{\mu}$ and solar $\nu_e$ can be explained
separately by considering each as a two neutrino system.  However,
both phenomena can be explained together if we consider a 3 neutrino
mass system. The relationship between the flavor and mass states of
the neutrinos can be expressed with the following matrix equation:

\begin{equation}
  \label{eqn:three-flavor}
  \left(
    \begin{array}{c}
      \nu_e  \\ \nu_{\mu} \\ \nu_{\tau} 
    \end{array} \right)
  = \left( \begin{array}{ccc}
      \rm{U}_{e1}     & \rm{U}_{e2}     & \rm{U}_{e3}    \\ 
      \rm{U}_{\mu 1}  & \rm{U}_{\mu 2}  & \rm{U}_{\mu 3} \\  
      \rm{U}_{\tau 1} & \rm{U}_{\tau 2} & \rm{U}_{\tau 3} 
    \end{array} \right)
  \left( \begin{array}{c}
      \nu_1  \\ \nu_2 \\ \nu_3 
    \end{array} \right) ,
\end{equation} 
\vspace{.05in}

\noindent
where the probability for a transition from flavor $a$ to flavor $b$
is given by:

\begin{eqnarray}
  P(\nu_a\rightarrow\nu_b) = 
  & \delta_{ab} & -4 \sum_{j>i}  {\rm Re}(\rm{U}^{\ast}_{ai}
  \rm{U}_{bi} \rm{U}_{aj}\rm{U}^{\ast}_{bj})\sin^2(1.27\Delta m_{ij}^2  L/E) \\ 
  & & \pm 2 \sum_{j>i}\rm{Im}(\rm{U}_{ai}^{\ast} \rm{U}_{bi}
  \rm{U}_{aj}U^{\ast}_{bj})\sin^2(2.54\Delta m_{ij}^2 L/E).
\label{eqn:transition-prob}
\end{eqnarray}

\noindent
with $L$ in km, $E$ in GeV, and $\Delta m^2$ in eV$^2$. The minus sign
refers to neutrinos and the plus sign to anti-neutrinos.  The familiar
two-flavor oscillation formula is a limiting case when considering
only a single $\Delta m^2$ between the two states.

With three neutrino masses, there are two neutrino mass differences
(\msqonetwo, \msqtwothree), three mixing angles (\tonethree,
\ttwothree, \tonetwo) and one CP violating phase. From our current
understanding of atmospheric and solar oscillations, we know the two
mass differences, and two of the mixing angles.  The $3 \times 3$
matrix from Eqn.~\ref{eqn:three-flavor} can be expressed in terms of
these angles as:

\begin{equation}
  \rm{U}= 
  \left(
    \begin{array}{ccc}
      1 & 0 & 0\\ 
      0 & c_{23} & s_{23} \\  
      0 & -s_{23} & c_{23} 
    \end{array} \right)
  \left(
    \begin{array}{ccc}
      c_{13} & 0 & s_{13}e^{i\delta}\\ 
      0 & 1 & 0 \\  
      -s_{13}e^{i\delta} & 0 & c_{13} 
    \end{array} \right)
  \left(
    \begin{array}{ccc}
      c_{12} & s_{12} & 0\\ 
      -s_{12} & c_{12} & 0 \\  
      0 & 0 & 1 
    \end{array} \right),
\end{equation} 
\vspace{.05in}

\noindent
where ``$s$'' represents the sine of each mixing angle and ``$c$''
represents the cosine.  In the decomposition above, the disappearance
of solar neutrinos is driven by the oscillations of the 1-2 mass
states, which are mixed by the matrix with the \tonetwo terms, and
observed atmospheric disappearance is driven by the matrix with the
\ttwothree terms.  The middle mixing matrix, contains the as yet
unmeasured \tonethree, which we hope to soon measure.  It should be
noted that the $\delta$ in this equation causes CP violation.  There
are two additional diagonal phases in the mixing matrix which adjust
the rate of neutrinoless double beta-decay if neutrinos are their own
anti-particles.  These phases do not effect oscillations as only
non-diagonal elements are present in equation~\ref{eqn:transition-prob}.

\subsection{Questions}

There are a set of remaining questions we hope to address in the set
of experiments which will be running in coming decade and beyond.  The
most important of these are:

\begin{enumerate}
\item What is the absolute mass of the neutrinos?
\item What is the pattern, or ``hierarchy'', of neutrino masses?
\item Are neutrinos their own anti-particles?
\item Is there CP violation in the neutrino sector?  If so, is it big
  enough to drive lepto-genesis and is it related to the quark sector?
\item What is the value of the final unmeasured mixing angle and is
  the atmospheric angle maximal?
\end{enumerate}

\section{ATMOSPHERIC AND SOLAR NEUTRINO EXPERIMENTS}

I would like to start by discussing recent results from atmospheric
and solar neutrino experiments.  Not only did these experiments give
us our first definitive evidence for neutrino oscillations but they
are continuing to add important information about neutrino
properties.  The best constraints on one of the mixing angles comes
from atmospheric neutrinos and the only measurement in a critical
region of oscillation transition comes from solar neutrinos.

\subsection{Atmospheric neutrino results}
\subsubsection{Super-Kamiokande}
The Super-K experiment has taken data in three phases and is now
starting a fourth phase to prepare for the upcoming T2K beam.  New
updated analyses were presented at this conference.  In preparing
these results the Super-K collaboration made improvements to the flux
calculations, reconstruction algorithms, Monte Carlo and detector
simulator, and also reevaluated their systematic errors.

These analyses exclude neutrino de-coherence and decay models at the
four and five sigma level, constrain oscillations into admixtures of
sterile neutrinos to be less than 23\%, and most crucially constrain
the atmospheric mixing angle \ttwothree to $45^\circ \pm 4^\circ$, a
10\% accuracy.  This is the most stringent current constraint on
\ttwothree.

\subsection{Solar neutrino results}
\subsubsection{Borexino}

\begin{figure}[!htbp]
  \centering
  \includegraphics[width=4.0in]{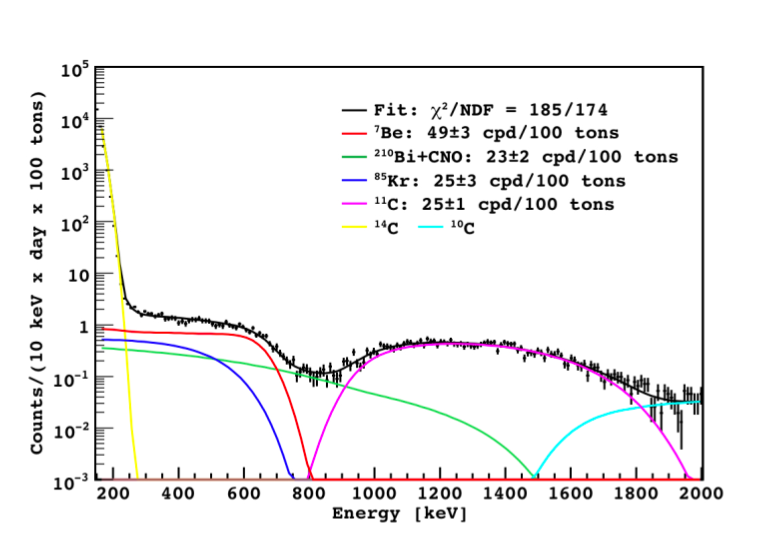}  
  \caption{The realtime observation of ${}^7Be$ solar neutrinos taken
    from~\cite{Arpesella:2008mt}. }
  \label{fig:borexino}
\end{figure}

\begin{figure}[!htbp]
  \centering
    \includegraphics[width=6.0in]{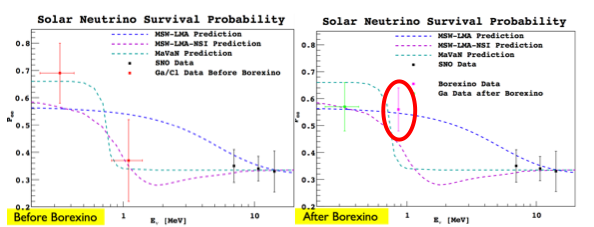}  
  \caption{Neutrino survival probability before and after Borexino.
    These figures were taken from Galbiati's Neutrino 2008
    presentation and annotated during my ICHEP08 talk.}
  \label{fig:before-and-after}
\end{figure}

One of the most important results in the last year came from the
Borexino experiment~\cite{Collaboration:2008mr}.  For the first time,
solar neutrinos in the energy region in the transition between vacuum
oscillations and matter-enhanced resonance oscillations have been
observed in real time. Previously, these low energy solar neutrinos in
the 1~MeV range had only been seen by radio-chemical experiments.  As
seen in Figure~\ref{fig:borexino} taken from~\cite{Arpesella:2008mt},

\noindent
a clear shoulder from the low-energy ${}^7Be$ neutrino scattering can
be seen on top of the low-energy radioactive background.  By using 192
days of data they were able to accurately measure the flux and show it
is consistent with the solar large mixing angle (LMA) solution.  This
is important because previously the large errors on the measurements in
this energy region from the gallium and chlorine experiments allowed
for exotic non-standard oscillation scenarios.  This is shown in
Figure~\ref{fig:before-and-after} where the effect of the Borexino
measurement can be seen.

Before Borexino, on the plot on the left, the two low energy points
are counting experiments using gallium and chlorine.  Since these are
integral measurements, the solar neutrino flux at any energy must be
calculated by subtracting the measured flux from higher energy
measurements from the measured flux.  When the higher energy chlorine
data is replaced by the Borexino measurement in the plot on the right,
the measurement from the gallium data also changes since the
contribution to the measurement from the higher energy flux is now
seen to be higher than previously assumed.  The effect of this is to now
have two data points with smaller error bars which fall nicely on the
LMA prediction.  A beautiful result!  This result directly confirms
our standard picture of neutrino oscillations using low-energy
neutrinos from the sun.

\subsubsection{Super-Kamiokande}

The search for direct evidence of LMA oscillations is also the goal of
the next phase of the Super-K solar analysis.  The Super-K
collaboration has worked hard during the SK-III phase to lower the
radioactive background in the central region of the detector.  If this
background rejection factor can be achieved in SK-IV along with lower
statistical errors and a lower correlated energy systematic error it
should be possible to see the turn-up in survival probability as
predicted by the LMA solution.  This should be on the order of a 10\%
effect in Super-K, and the plan for SK-IV is to lower the energy
threshold to 4.0~MeV to make this measurement possible.

\subsubsection{SNO Phase III}

The SNO experiment was the experiment that first conclusively
demonstrated that the deficit of solar neutrinos seen on the earth was
also due to neutrino oscillations. They did this by measuring both the
flavor-blind neutral-current reaction, and the electron-neutrino only
charged current reaction off of deuterium. Although the number of
measured charged current reactions was less than expectation, the
number of neutral current reactions was as expected, showing that the
electron neutrinos were not disappearing but rather just transforming
into a flavor for which the detector had no sensitivity.

Doing this measurement requires tagging the neutron that is produced
in the neutral current reaction:

\begin{equation}
  \label{eq:1}
  \nu_x + d \rightarrow p + n + \nu_x.
\end{equation}

There were three ways to see this neutron and they corresponded to the
three running phases of SNO.  In the first phase the neutron was
captured on pure $D_2O$.  In the second phase, salt was added to the
water increasing the neutron capture cross-section and resulting in
the production of higher energy gamma-rays which were easier to
detect.  In the third and final phase, for which we saw results at
this conference, ${}^3 \rm He$ neutron counters were added to the detector.
Where as before, the energy from both the charge and neutral current
reactions was measured via Cherenkov light, now the NC signal was
measured in the neutron counters thereby breaking a degeneracy of the
systematic errors and adding confidence to the measurements.

\subsubsection{Kamland 2008}

\begin{figure}[!htbp]
  \centering
  \includegraphics[width=4.0in]{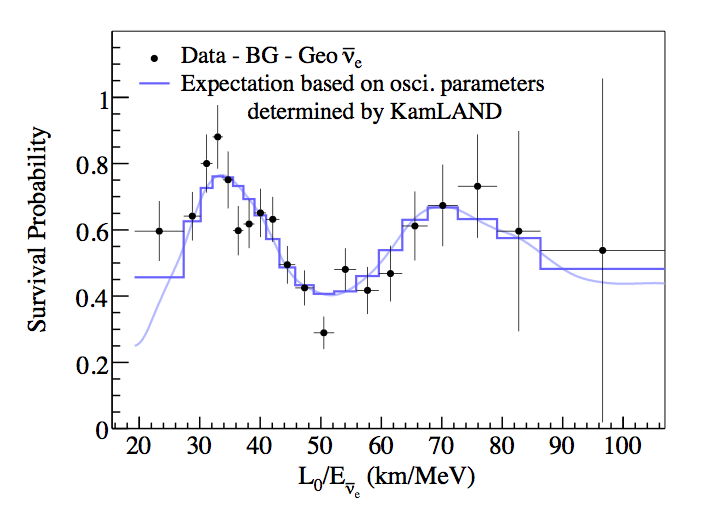}    
  \caption{New results from the Kamland experiment. This figure is
    taken from~\cite{:2008ee} and shows almost two full cycles of
    oscillation.}
  \label{fig:kamland}
\end{figure}

Finally, I want to mention the recent results from the Kamland
experiment which probe the same oscillation parameters as the solar
experiments but use neutrinos from reactors.  This year, we saw
exciting new results from Kamland where they increased the size of
their data set and also, by improving their analysis techniques, were
able to both lower the energy threshold to 1~MeV and use data closer
to the wall. The combined effect of this was an increase of almost a
factor of 20 in exposure.  The result can be seen in
Figure~\ref{fig:kamland} taken from~\cite{:2008ee} which shows a clear
sinusoidal oscillation pattern over almost two full cycles!

Covering such a large range in L/E means that the fit to the mass
squared difference can be very precise and these results allow Kamland
to measure the solar mass splitting to the level of 2.7\%.

\section{OSCILLATION SEARCHES AT ACCELERATORS}

I would now like to turn to searches for neutrino oscillation using
accelerators. Before I do, I would like to explain the two main
classes of accelerator searches.  There are two types of searches that
can be undertaken at accelerators. There are both {\it appearance} and
{\it disappearance} searches. The differences comes down to whether or
not the neutrinos produced in the beam that is being used have enough
energy to produce their associated lepton via a charged-current reaction.  
For example,  if a muon neutrino has 2 GeV of energy, there is no
problem to produce a muon when it interacts off of a nucleus.  If, on
the other hand, a muon neutrino has only 5~MeV of energy there isn't
enough energy available to produce a muon.  

So, if we were to start with an electron neutrino which later
oscillated into a muon neutrino, in the first case there would be no
problem creating the muon and we could (for example) look for muons to
appear from our electron beam.  In the second case however, the charged
current reaction couldn't happen and the neutrinos will have seemed to
have disappeared.  It should be noted that in both of these cases the
neutral current reactions are unaffected since the neutral current
interactions are flavor blind.  Examples of appearance analyses and
experiments that I am presenting today include the Miniboone, T2K,
No$\nu$a and OPERA experiments.  Examples of disappearance experiments
include MINOS, K2K, and the Super-K atmospheric analysis.

\subsection{MiniBooNE Results}

In the beginning of this talk I pointed out that there appear to be
three neutrinos and two mass differences.  Until very recently however,
there was a piece of evidence which challenged this picture.  The LSND
experiment had presented evidence of a third mass
splitting~\cite{Athanassopoulos:1996jb} when they saw evidence for
\nue appearance from a muon beam coming from stopped pions.  If this
result was true it would mean either that there were more neutrinos
than we knew about (which is hard to accommodate with other data) or
that our ``simple'' picture of neutrino oscillation was too naive.
So, a new experiment, MiniBooNE was build to test the same region of
L/E as the LSND experiment but at a different baseline and energy so
as to have difference systematic errors.

Approximately a year ago, the MiniBooNE collaboration reported it's
first results~\cite{AguilarArevalo:2007it} reporting that there was no
appearance signal where one would expect if the LSND result was correct
in the model of ``simple'' two or three flavor oscillations.
Puzzlingly however, an unexpected excess of data over Monte Carlo was
seen in the lowest energy part of their sample below where the LSND
signal was expected.

The collaboration spent this last year intensively studying this
anomaly.  They included more data in their search, included new
backgrounds in their analysis, and re-evaluated the backgrounds they
had already included.  We saw those results for the first time in this
meeting.  The collaboration concluded that the excess was still
significant.  The results of their oscillation search remains
unchanged, they still rule out the previous LSND results in a standard
two flavor oscillation scenario but they cannot yet explain the
low-energy excess.

The collaboration is undertaking some important (and impressive!)
cross-checks now.  First of all, we will hopefully soon see results
from their anti-neutrino running which should shed light on whether
this is a cross-section effect (and even more directly test the LSND
result which was performed with anti-neutrinos).  Secondly, and most
impressively they are also looking at highly off-axis neutrinos from
the NuMi neutrino beam used by the MINOS experiment.  This beam is at
a very different energy and baseline then their beam so if this effect
is an oscillation effect it should appear at a different energy than
what is being seen now.

\subsection{MINOS Results}

The MINOS collaboration uses the muon neutrino beam produced at the
NuMi facility at Fermilab to shoot neutrinos over 700~km away to their
detector in northern Minnesota.  MINOS has now measured the
\msqtwothree parameter to approximately 5\%, the best measurement on
the parameter to date.  They have also excluded decay and decoherence
models at the 4-6 sigma level and ruled out oscillation into
admixtures sterile states to a fraction of less than 68\%.  MINOS will
continue to take data resulting in a further tightening on the
measurement of \msqtwothree along with a chance to measure \tonethree
through \nue appearance if the value of the parameter is close to the
current limit.

\section{FUTURE ACCELERATOR OSCILLATION PROSPECTS}

There are three accelerator based oscillation experiments that are
either starting or planning on starting soon.  All three look for the
appearance of a neutrino flavor that was not in the original beam.

\subsubsection{OPERA}

The first of these experiments is the OPERA experiment which has
already begun operation.  The OPERA experiment aims to see the direct
appearance of tau neutrinos in a beam of muon neutrinos produced over
700~km away at CERN.  They hope to do this by looking for the
tell-tale kink of the tau decay in a emulsion block.  

OPERA is a hybrid electronic/emulsion detector.  Electronic trackers
including drift chambers point tracks back to emulsion blocks which
are removed by robots and scanned by a network of automatic scanning
stations.  After 1.5~years, the installation of the blocks is now
complete.  Depending on the exact value of the atmospheric mixing parameter
the experiment expects to measure on the order of 10 events on a
background of less than 1 event after 5 years of running.  The full
beam is expected in 2009 and the experiment is on track to see their
first tau interaction next year.

\subsubsection{T2K}

The T2K experiment is one of two new so-called ``off-axis''
experiments.  Unlike K2K and then MINOS, the neutrino beam is aimed a
few degrees away from the far detector.  By aiming the beam off-axis
the far detector sees fewer total neutrinos but a sharper narrow band
beam with more neutrinos at the energy of interest.  Due to the
kinematics of pion decay, most of the neutrinos at a particular angle
from the beam direction give the same neutrino energy.  This allows the
experiments to probe neutrino appearance and disappearance at the
particular energy that is predicted by the oscillation model.

The main goal of the T2K experiment is to observe the appearance of
\nue in a almost pure \numu beam after traveling the 295~km
between the JPARC accelerator center and the Super-K detector.
Detection of electron appearance would allow a measurement of the
unknown \tonethree mixing angle.  The expected sensitivity is ten to
twenty times better than the current limit from the CHOOZ
experiment~\cite{Apollonio:2002gd}.  Additionally, the T2K experiment
hopes to measure the two atmospheric mixing parameters \msqtwothree
and \ttwothree to a few percent accuracy hopefully determining whether
or not the atmospheric mixing angle is maximal.

T2K employs both on-axis and off-axis near detectors located 280~m
from the neutrino source along with the Super-Kamiokande detector
295~km away.  The off-axis near detector uses the refurbished UA1
magnet along with FGDs, TPCs, and a water/scintillator detector for
measuring the neutrino interactions before the neutrinos have a chance
to oscillate.

The T2K neutrino beam will start in April of 2009 and the on-axis
detectors along with Super-K will be operational at that time.  In the
fall of 2009 the off-axis detector will be installed and full running
will begin in the winter.

\subsubsection{No$\nu$a}

The No$\nu$a experiment is also an off-axis experiment optimized for
the search for \nue appearance.  No$\nu$a will make use of a 700~kW 
upgraded version of the already existing NuMi beam-line.  The detector
will be located over 700~km away in northern Minnesota at Ash River
near the Canadian border.  

The No$\nu$a far detector is a 15~kton totally active liquid
scintillator detector.  It, and its matching off-axis near detector,
function as a low-Z calorimeter allowing time for electromagnetic
showers to develop over a long distance to help distinguish showers
initiated by electrons or pizero decays. No$\nu$a is unique in the
current generation of long-baseline experiments in that its ~700~km
long baseline will allow it to address question of the mass hierarchy
if the value of \tonethree is large enough.  It will be do this by
comparing the \numu to \nue oscillation probability for both neutrino
and anti-neutrino beams.

The No$\nu$a collaboration plans to start construction of their far
detector in April of 2009, with the first 2.5~kton of data taking
in August of 2012.  The detector should be complete in January of
2014.  Like T2K it has an expected sensitivity to \tonethree of 10 to
20 times lower than the present limit.

\section{$\Theta_{13}$ AT REACTORS}

There is another complimentary way to measure \tonethree by using
neutrinos from reactors.  Unlike the accelerator based experiments,
these are disappearance experiments and use the same inverse beta
decay capture reaction as used by the Kamland experiment.  Neutrinos
from a powerful reactor are measured just a few kilometers away from
the cores and monitored for a distortion in the energy spectrum.  This
technique is complimentary to the accelerator experiments since they
are measuring slightly different things.  Unlike the long baseline
appearance probability the disappearance probability has no dependence
on the atmospheric mixing angle or CP violating phase.  In the first
few kilometers from the reactor core the disappearance probability is
given by:

\begin{equation}
  \label{eq:2}
  P (\bar \nu_e \rightarrow x) \approx \sin^2 \theta_{13} \sin^2
  \left(  { \Delta m^2_{31} L \over 4 E}   \right).
\end{equation}

This is the technique that was by the CHOOZ
collaboration~\cite{Apollonio:2002gd} when it set the current best
limit on the \tonethree parameter.  Several collaborations hope to
redo this measurement with a much tighter control of the systematic
errors, allowing them to achieve substantially lower sensitivity than
before.  Current and planned experiments include the Double CHOOZ
experiment in France ( 8.7 GW reactor), Daya Bay in China ( 11.6 GW),
Reno in Korea (17.3 GW) and Angra (6 GW) in Brazil.

\subsubsection{Double CHOOZ}

In order to achieve a much lower sensitivity than the previous version
of the experiment, the Double CHOOZ experiment will reuse the previous
CHOOZ site 1051~m from the reactor core but add a second site 300m
from the reactor core.  By using two identical detectors the first
detector will be used to normalize the results in the farther detector
much as in the accelerator experiments.  

The key to these new class of reactor experiments is to rely on
relative measurements.  By also building more robust veto and
calibration systems they can lower the expected backgrounds.  They
hope to decrease the statistical error on the experiment from 2.7\% in
CHOOZ to be .5\% in the new configuration.  

The far lab is being built now and they plan to have the far detector
running in early 2009.  The near detector lab construction should
begin in late 2008 with assembly of the near lab completed in late
2009.  After approximately 1.5 years of running the Double CHOOZ
experiment should achieve a sensitivity on $\sin^2 2 \theta_{13}$ of
approximately 0.6. Then, the near detector should become operational
reaching a sensitivity of 0.03 after a few more years or running.

\subsubsection{Daya Bay}

The Daya Bay collaboration has the impressive goal of measuring
$\sin^2 2 \theta_{13}$ to the 0.01 level, similar to the accelerator
based experiments.  In order to achieve this difficult task they need
to keep both the statistical and systematic errors very low.  To
achieve this they are taking a multi-prong approach.

First off, they are optimizing the distance of their detectors to
maximize the oscillation signal and they are placing the detectors as
deep as possible to try to reduce cosmogenic backgrounds. To further
reduce backgrounds they are making a multi-level comprehensive veto.
They will leverage their high power reactor with a large detector mass
spread over multiple near and far detectors.  They plan for 160~tons of
target distributed in 8 detectors and may swap detectors between the positions.

The surface building assembly began in summer of 2008 and they expect
data running in all eight detectors in three halls by December 2010.
If all goes according to plan they will reach a sensitivity of 0.01 in
2013 or 2014.

\section{DIRECT MASS MEASUREMENTS}

All of my talk up-to-now has been concerned with oscillation studies.
These studies probe the mass splittings and mixing angles but tell us
nothing about the absolute mass scale of neutrinos.  Along with
cosmological techniques, which I will not cover here, there are two
main techniques for studying the absolute neutrino mass.

\subsection{KATRIN}

The first technique is the direct measurement of kinematic parameters
in beta decays.  This is an old idea and the two experiments which
have the best current limits on the absolute mass of the neutrino
(Mainz~\cite{Kraus:2004zw} and Troisk~\cite{Belesev:1995sb}) have
come together to form a new collaboration known as Katrin.

In direct mass experiments, a radioactive beta source such as ${}^3
\rm He$ is used and the energy spectrum of the outgoing electron is
examined.  If the neutrino has mass, then the end point of the
spectrum will be distorted.  The previous limits are on the order of
2~eV and Katrin hopes to push them to 0.2~eV.  In order to be
sensitive to distortions in the end point in the last eV you need both
very high statistics and good energy resolution.

High statistics are necessary because only $2 \times 10^{-13}$ of all
decays take place in the last 1~eV.  So, in order to measure this
shape you need a huge number of decays.  Energy resolution is
important because the small distortion in the end-point energy will be
washed out if the energy resolution is not less than this difference.
Katrin addresses these two issues with an impressive 70~m detector.  A
high intensity tritium source produces electrons which are transported
first through a pre-spectrometer and then into a large volume
spectrometer which acts as a high-pass filter with an adjustable
threshold.  The electrons which pass through the spectrometer are then
brought to impinge on a multi-pixel silicon semiconductor detector
which has very high energy resolution and a very thin entrance window.

\subsection{Double Beta Decay}

There is another approach to measuring absolute neutrino mass that has
the added advantage that is has the potential to also tell us
something about the mass hierarchy and the Majorana nature of the
neutrino.  In some nuclei, normal beta-decay can't happen because it
is energetically disfavored.  In these cases, it is still possible for
two beta decays to happen simultaneously.  This process is known as
double beta-decay and happens at a much lower rate then normal
beta-decay.  When double beta-decay takes place two electrons and two
neutrinos are emitted and if one looks at the sum the energies of the
electrons you get a beta-decay like spectrum.

If, however, neutrinos are their own anti-particles (Majorana
particles) then something special can happen.  The two neutrinos can
annihilate with each-other in a process known as neutrino{\it less} double
beta-decay (as opposed to the normal neutrino{\it full} double
beta-decay. When this happens the kinematics of the outgoing particles
are constrained and the sum of the two electron energies is no longer
continuous but rather monochromatic.  In order to see neutrinoless
double beta-decay one must find the monochromatic line at the end of
the continuous neutrinofull double beta-decay spectrum.

If this process can be measured, it tells you a lot.  First of all, it
tells you that neutrinos are Majorana, that they are their own
anti-particles.  This has very serious implications for models of
neutrino mass generation. Secondly, it tells you about the absolute
mass of the neutrinos.  What you actually measure in double beta-decay
is a rate of the double beta-decay process.  This rate is proportional
to three important quantities: a phase space factor, a nuclear matrix
element, and the effective mass.  The effective mass can be written as:

\begin{equation}
  \label{eq:3}
  <m_{\beta \beta}> = \left| \Sigma_i U_{ei}^2 m_i \right| .
\end{equation}

Where the $U_{ij}$ are the elements of the mixing matrix in
Equation~\ref{eqn:three-flavor} and include two diagonal Majorana
phases.  The electron coupling is mostly to the first two mass states
and so the hierarchy is extremely important.  In an inverted hierarchy
the first two mass states have a minimum mass of the atmospheric
neutrino splitting while in the normal hierarchy the lowest mast state
could have an absolute mass as low as zero.  This is demonstrated in
the following figure taken from~\cite{Avignone:2007fu} which shows the
effective mass verses the minimum mass of the lowest mass state.

\begin{figure}[!htbp]
  \centering
  \includegraphics[width=4.0in]{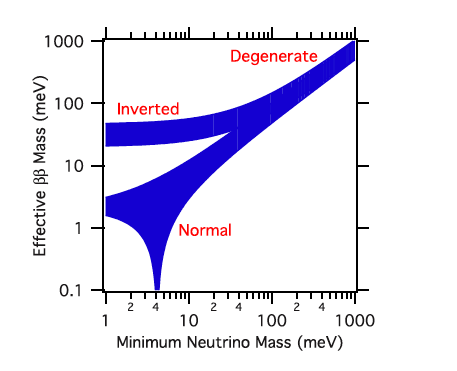}    
  \caption{The effective mass verses the mass of the lowest mass
    state. Figure is taken from~~\cite{Avignone:2007fu}. }
  \label{fig:effective-mass}
\end{figure}

For the current and upcoming generation of experiments we can hope to
probe the 10s to 100s of meV range of mass if we have an inverted
hierarchy.  If however, the hierarchy is normal we will have to wait
for even larger experiments.

\subsubsection{The relationship to accelerator experiments}

Non-accelerator and accelerator based neutrino experiments are often
thought of separately but there are very important connections between
them.  The case of the mass hierarchy is an important example.  If
neutrinoless double beta decay is seen in this next generation of
experiments we will know that neutrinos are Majorana and that we are
dealing with an inverted hierarchy.  This in turn tell us we should
expect to measure an inverted hierarchy in long-baseline experiments.
If we didn't, then there would be an important flaw in our theory.
Conversely, if we measure the normal hierarchy in our long-baseline
experiments we should not expect to be able to measure neutrinoless
double beta-decay in the near future.

What are we to think if absent evidence from long-baseline experiments
we measure nothing in the upcoming round of double beta-decay
experiments?  It could mean several things:  Neutrinos might be Dirac
particles which would preclude the process from taking place, we might
be dealing with a normal hierarchy, the mass might be just below our
reach, or most depressingly, we might be unlucky enough that the
Majorana phases have conspired to cancel out and reduce our expected
rate to near zero.

\subsection{Experimental Techniques}

There are very many collaborations active in this field and many
proposed experimental techniques.  I will just touch on a few here
that presented talks at this conference and serve as examples of some
of the major classes of experiments.

\subsubsection{Gerda/ Majorana}

The Gerda and Majorana collaborations both use solid state germanium
detectors.  They are using enriched Ge-76 diodes which have excellent
resolution. They are using somewhat different techniques,
Gerda suspends bare Ge in a cryogenic bath, while Majorana uses a more
traditional cryostat approach.  The two collaborations are working
together to characterize detectors and build common Monte Carlo tools.
The Gerda collaboration plans on finishing construction of their
phase-I experiment by spring of 2009.

\subsubsection{EXO}

The EXO experiment uses cryogenic liquid xenon and measures both the
scintillation light and charge in a TPC.  The gas is easily repurified
and recycled and heavy R\&D is taking place for future extremely large
volumes which would attempt to tag the remaining Ba daughter isotope
so as to remove background.  A 200~kg prototype has been installed in
the WIPP underground facility and cool-down is expected to be complete
in early 2009.  

\subsubsection{SNO+}

The SNO+ experiment aims to build a very large self shielding mass of
scintillator for the purposes of double beta-decay searches in
addition to solar, reactor, geo and supernova neutrino studies.  They
hope to do this be refilling the old SNO vessel with scintillator.  By
doping the scintillator with ${}^{150} \rm Nd$ which has an endpoint energy
above most known low energy backgrounds, they hope to fit the endpoint
spectrum plus neutrinoless beta-decay signal. Now they are doing R\&D
to determine the best way to dope the scintillator and are working to
fill the vessel soon.

\section{SUMMARY AND FUTURE}

I hope this talk has given you an overview of the experimental
neutrino picture. Putting all of the information together, the
following two figures taken from~\cite{:2008ee} and prepared by
E. Kearns respectively show the current state of the art for measuring
the solar and atmospheric oscillation parameters.

\begin{figure}[!htbp]
  \centering
  \includegraphics[width=4.0in]{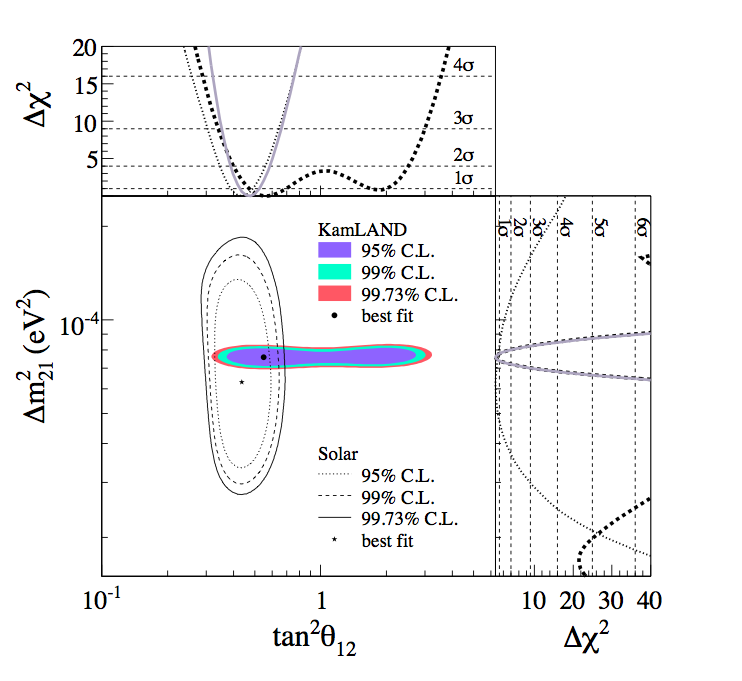}
  \caption{The solar mixing parameters as measured by the Kamland and
    and solar neutrino experiments.  This figure taken
    from~\cite{:2008ee}.}
  \label{fig:solar-param}
\end{figure}

\begin{figure}[!htbp]
  \centering
  \includegraphics[width=4.0in]{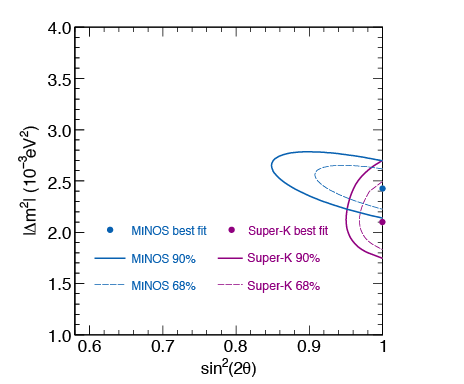}
  \caption{The atmospheric mixing parameters as most recently measured
    by the Super-K and MINOS collaborations.  This figure was prepared
    by E. Kearns.}
  \label{fig:atmos-param}
\end{figure}

We now have a consistent set of experiments with measurements at the 2
- 10\% level. The best measurements of \msqonetwo come from Kamland,
on \tonetwo from the solar experiments, on \msqtwothree from MINOS and
on \ttwothree from Super-K.  They all point to the same places in
parameter space.  The new results from Borexino and MiniBoone further
cement our understanding of the situation.

The next set of experiments will hopefully get us close to the 1\%
level in accuracy.  The period of precision neutrino experiments has
begun.  We also hope to add info on absolute mass of the neutrino to
the picture along with the nature of the mass pattern and whether the
neutrino is a Dirac or Majorana particle.  All of the experiments I
mentioned above must be considered together for a complete
understanding of the situation.

\subsection{Future Facilities}

It is also worth spending a moment to discuss even farther future
ideas and plans.  Many groups in the US, EU and Asia are thinking
about how to measure the CP violating phase $\delta$ if \tonethree is
seen in the next generation of experiments.  The technique for
measuring $\delta$ is to compare the rate of neutrino and
anti-neutrino oscillation.  Unfortunately, it turns out that this
measurement will only give measurable results of \tonethree is large
enough, which is why the next generation of long-baseline experiments
are so critical to this question.  

There is more than one approach to producing the neutrino and
anti-neutrino beams used in these studies.  First of all, there are
the so called ``super-beams''.  These are more powerful versions of the
conventional pion decay based beams we are utilizing in long-baseline
experiments now.  Next there are beta-beams.  These beams of very pure
electron neutrino or anti-neutrino beams are made by allowing
accelerated radioactive ions to decay in a storage ring.  Finally,
there are neutrino factories.  In a neutrino factory muons are
produced from pion decay, cooled, injected into a storage ring and
allowed to decay in long straight sections.  

When possible,  future facilities should be incorporated into a larger
program of particle physics.  A nice example of this idea is the
proposed Project-X facility at Fermilab.  Project-X is a super-beam
which would send a very high intensity conventional neutrino beam to the
proposed Deep Underground Science and Engineering Laboratory in South
Dakota.  This facility would also provide a high intensity muon source
for other lepton physics and could be the first stage of a future
neutrino factory or muon collider.  

Finally, I want to mention that there is a large world-wide R\&D
program taking place on future large detectors to serve as targets for
these beams.  The work now is on designing extremely large water
Cherenkov and liquid argon detectors.
 
\subsection{Things I skipped}

In covering such a large field there are many items I regrettably had
to skip in this talk.  I have prepared backup information for some of
them which you can find in my slides.  In this conference we heard
about future experimental atmospheric neutrino prospects from the INO
experiment in India which will utilize RPCs and a magnetic field to
measure first atmospheric neutrinos, and then hopefully neutrinos from
a neutrino factory.

We also had nice reviews and results on both hadron production
experiments and neutrino interactions. Both of these topics are vital
to understanding precision neutrino oscillation results from
long-baseline experiments.   Finally, we heard nice results on low
energy scattering experiments and saw a beautiful set of measurements on
normal two neutrino beta decay from the NEMO-III experiment along with
their plans for the Super-Nemo facility.  

There is a large list of other topics which I might have covered if I
had time and I would have particularly liked to have said something
about supernova and how the neutrinos measured on Earth from one could
tell us about particle physics and neutrino properties in addition to
astrophysics.

\section{CONCLUSIONS}

The last ten years have seen the discovery of neutrino mass, the
solution to the solar neutrino problem and the start of precision
neutrino measurements.  So far, everything fits together well.  If we
are lucky, in the next decade we will measure the remaining mixing
angle, learn the nature and absolute mass of the neutrino along with
its mass pattern.  This information coupled with new knowledge gained
at the LHC will let us plan our next steps.

What I outlined above is an ambitious program that will take many
groups working in concert over many years to achieve.  Preparing for
this talk let me see that there are indeed large numbers of excited
people working on these problems. Because of this I am quite hopeful
that the person who is lucky enough to talk to you ten years from now
will have as many good results to report to you as I did today.

On this note I'd like to close with the end of a poem written by
B. Nolty and K. Scholberg from our days together as graduate students
on the MACRO experiment:

\begin{center}
  {\it
    ``So gladly we march to mines, tunnels or seas: \\
    New physics or not, we get PhDs -- \\
    Condensed to a preprint we'll mail to our mothers \\
    Of one page of physics (two pages of authors).  }
\end{center}
\noindent
{\it (Full text at http://www.cithep.caltech.edu/macro/songs/sad.html)}


\end{document}